\documentclass[journal]{IEEEtran}
\usepackage[ruled,vlined]{algorithm2e}
\SetAlgoSkip{6pt}
\SetAlgoInsideSkip{1pt}
\usepackage{cuted}
\usepackage{mathrsfs}

\usepackage{algcompatible}
\usepackage{algpseudocode}
\algnewcommand\algorithmicinput{\textbf{Input:}}
\algnewcommand\INPUT{\item[\algorithmicinput]}
\algrenewcommand\algorithmicindent{1.2em}
\usepackage{cite}
\usepackage[subtle]{savetrees}
\usepackage{verbatim}
\usepackage{graphicx}
\usepackage{siunitx}
\usepackage{amsmath,amssymb,amsfonts,bm}
\usepackage{epsf}
\usepackage{tikz}
\usepackage{psfrag}
\usepackage{pstool}
\usepackage{epstopdf}
\usepackage{bbm}
\usepackage{booktabs}
\usepackage{stfloats}
\usepackage{url}
\usepackage[dvips]{epsfig}
\usepackage{amsthm}
\usepackage[usenames,dvipsnames]{pstricks}
\usepackage{lettrine}
\usepackage[dvips]{epsfig}
\usepackage{pst-grad}
\usepackage{pst-plot}
\usepackage{epsfig}
\usepackage{enumerate}
\usepackage{amsbsy}
\usepackage{amssymb}
\usepackage{amsthm}
\usepackage{amscd}
\usepackage{float}
\usepackage[caption = false]{subfig}
\usepackage{stackrel}
\usepackage{authblk}
\usepackage{dsfont}
\usepackage[normalem]{ulem}
\usepackage{hyperref}
\usepackage{xcolor}
\usepackage{enumitem}

\makeatletter
\newcommand{\removelatexerror}{\let\@latex@error\@gobble}
\makeatother

\newtheorem{prop}{Proposition}

\usepackage{stackengine}

\definecolor{mintbg}{rgb}{.63,.79,.95}
\setlist[itemize]{leftmargin=*,nosep}

\begin{document}


\title{{Risk-Aware Joint Inference and Resource Control for Physical AI in 6G Networks}}

\author{Amin Farajzadeh, \textit{Member}, \textit{IEEE}, Melike Erol-Kantarci, \textit{Fellow}, \textit{IEEE}
\thanks{A. Farajzadeh and M. Erol-Kantarci are with the NETCORE Lab, School of Electrical Engineering and Computer Science, University of Ottawa, Ottawa, ON K1N 6N5, Canada (e-mails: \{amin.farajzadeh, melike.erolkantarci\}@uottawa.ca).}}
\pagenumbering{gobble}
\makeatletter
\patchcmd{\@maketitle}
  {\addvspace{0.5\baselineskip}\egroup}
  {\addvspace{-1.8\baselineskip}\egroup}
  {}
  {}
\makeatother

\maketitle

\begin{abstract}
Wireless edge inference enables Physical-AI devices to execute edge-inferred
actions using transmitted sensory inputs. Since the risk of incorrect actions vary across observations and tasks, the associated uplink transmission of sensory inputs requires different levels of wireless reliability. For instance, applying a fixed signal-to-interference-plus-noise ratio (SINR) requirement can overprovision low-risk tasks while underprovisioning high-risk ones. Therefore, we propose risk-aware control for edge inference (RACE), which maps predicted action-failure probability and task criticality to adaptive SINR targets and combines them with deadline-derived SINR requirements. A slow-timescale large language model (LLM) proposes task requirements for deterministic validation, while a fast hybrid actor selects inference configurations and transmit power levels, complemented by analytical receive beamforming and finite power restoration. Simulations show $11\%$ improvement in the average accuracy--latency--energy objective over fixed-high-SINR control, with near-zero constraint outage.
\end{abstract}
\vspace{-1mm}

\IEEEpeerreviewmaketitle

\vspace{-1mm}
\section{Introduction} \label{sec:introduction}
Sixth-generation (6G) networks are expected to support distributed AI inference for robots and other Physical-AI devices \cite{thomas2026holonic,kumbhar2026physicalai}. In wireless edge inference, where limited device computational resources may hinder on-device inference, a device locally extracts and encodes task-relevant features from its sensory observations, transmits them to an edge server for inference, and executes the returned physical action~\cite{li2024semanticarq}. Wireless communication therefore affects the timely and reliable delivery of information needed for physical decisions. Supporting these tasks requires balancing inference accuracy, end-to-end latency, and communication and computation energy~\cite{yuan2026featurecompression}.

Existing task-oriented wireless and edge-inference studies have jointly optimized feature transmission, inference performance, and radio resources \cite{shi2023taskoriented6g,wen2024taskorientededgeai, zhang2024cloudran,xie2026hybridprecision,li2026taskprecoding}. However, average task performance alone does not capture how the risk of an incorrect physical action varies across observations and operating conditions. 
Consider a mobile robot navigating an empty corridor versus approaching a partially occluded person. In the latter case, the observation may be harder to interpret and incorrect actions may have more serious consequences. Although both cases require wireless feature transmission, their differing risks call for wireless resource control that accounts for both the likelihood and consequences of inference errors.

Motivated by this challenge, we propose risk-aware control for edge inference (RACE), a two-timescale framework that jointly controls inference configurations and wireless resources according to physical-action risk and task deadlines. At the slow timescale, a large language model (LLM) uses task descriptions and feedback to propose two task parameters: physical criticality, representing the consequence of an incorrect action, and the maximum allowable end-to-end action delay. Unlike LLM-assisted semantic encoding and decoding in~\cite{salehi2025llm}, the LLM here proposes task requirements rather than processing transmitted features. A deterministic validator checks criticality bounds and deadline feasibility, accounting for inference and feedback delays, and applies a predefined fallback to invalid proposals. The resulting task contract guides fast-timescale control.

At the fast timescale, a local predictor estimates, before each uplink transmission, the probability of an unacceptable action error under a fixed reference inference configuration. Combined with the validated task criticality, this yields a task-risk metric, which determines an adaptive signal-to-interference-plus-noise ratio (SINR) target. The task deadline imposes an additional, configuration-dependent SINR requirement based on the feature size and available transmission time; the larger of the two defines the effective SINR target. A hybrid actor then selects inference configurations and transmit power levels for all devices, while the base station computes receive beamformers analytically and applies finite power restoration to improve SINR constraint satisfaction. RACE thus seeks to minimize inference-related action loss, end-to-end latency, and energy consumption under SINR, deadline, and power constraints, without involving the LLM in time-critical wireless decisions.

To the best of our knowledge, RACE is the first framework to incorporate both predicted physical-action failure probability and task-specific consequences into risk-aware wireless resource allocation decisions for edge inference in Physical-AI networks. The main contributions are summarized as follows: 
\begin{itemize}[leftmargin=*]
\item We formulate risk-aware wireless edge inference by combining predicted action-failure probability and physical-action consequence into a risk metric that directly determines adaptive SINR requirements. 
\item We develop RACE, combining validated slow-timescale task requirements with fast hybrid actor--critic control of inference configurations and transmit power levels, analytical receive beamforming, and finite power restoration. 
\item Simulations demonstrate a $11\%$ improvement in average accuracy--latency--energy objective relative to the reliable fixed-high-SINR baseline, with near-zero constraint outage.
\end{itemize}
\vspace{-3mm}
\section{System Model and Problem Formulation}
\label{sec:system_model}

We consider a Physical-AI-native wireless edge network where an $M$-antenna base station (BS) with an edge inference server supports a set $\mathcal K=\{1,\ldots,K\}$ of embodied devices. To reduce device computational demands, all physical-action inference is offloaded to the edge server, excluding on-device inference. Each device locally extracts a task-relevant sensory representation and relies on uplink feature transmission, BS receive beamforming, and edge inference to obtain a physical-action decision. Time is divided into slow-timescale task intervals indexed by $n$, where interval $n$ contains a set $\mathcal T_n$ of wireless slots.

At slot $t\in\mathcal T_n$, device $k$ observes a sensory input $\mathbf X_k(t)\in\mathcal X_k$ generated by its physical environment. Let $\mathbf Y_k(t)\in\mathcal A_k$ denote the ideal physical action associated with that input, where $\mathcal A_k\subseteq\mathbb R^{m_k}$ is the device action space. The pair $(\mathbf X_k(t),\mathbf Y_k(t))$ follows the unknown task distribution $\mathbb P_k^{(n)}$ over $\mathcal X_k\times\mathcal A_k$. During online operation, device $k$ observes only $\mathbf x_k(t)$, whereas $\mathbf Y_k(t)$ is available only in labeled offline data. 

A fixed local feature extractor
$f_k^{\rm loc}:\mathcal X_k\rightarrow\mathbb R^{d_k}$ maps the sensory
input to $\mathbf z_k(t)=f_k^{\rm loc}(\mathbf x_k(t))$. For device $k$, let
$\mathcal C_k=\{1,\ldots,N_k\}$ denote a finite library of pretrained and profiled feature/inference configurations. Configuration
$c\in\mathcal C_k$ specifies
$(
        e_{k,c},
        g_{k,c},
        S_{k,c},
        T_{k,c}^{\rm inf},
        P_{k,c}^{\rm inf}
    ),$
where
$e_{k,c}:\mathbb R^{d_k}\rightarrow\{0,1\}^{S_{k,c}}$
is a feature encoder,
$
g_{k,c}:\{0,1\}^{S_{k,c}}\rightarrow\mathcal A_k
$
is the corresponding edge-side decoding and inference mapping,
$S_{k,c}$ is the encoded feature length in bits,
$T_{k,c}^{\rm inf}$ is the measured edge execution time, and
$P_{k,c}^{\rm inf}$ is the average edge processing power. 
Thus, for the selected configuration $c_k(t)$, the transmitted bit vector and
edge-generated physical action are
\vspace{-2mm}
\begin{equation}
    \mathbf b_k(t)
    =
    e_{k,c_k(t)}(\mathbf z_k(t)),
    \qquad
    \widehat{\mathbf Y}_k(t)
=
g_{k,c_k(t)}(\mathbf b_k(t)).
    \label{eq:configured_edge_inference}
    \vspace{-2mm}
\end{equation}
The inferred action $\widehat{\mathbf Y}_k(t)$ is returned to device $k$ and
applied after the feedback delay $T_k^{\rm fb}$. The quantities
$S_{k,c}$, $T_{k,c}^{\rm inf}$, and $P_{k,c}^{\rm inf}$ are obtained offline
for the deployed device--server platform.

After extracting $\mathbf z_k(t)$ and before encoding the payload, device $k$ evaluates and reports its pre-inference uncertainty to the BS. Based on these uncertainty values and the available channel state information (CSI), the BS selects $c_k(t)$ and transmit power $p_k(t)$ and computes the receive beamformers. The selected configuration and power are signaled through a low-rate control link. Feature-extraction and control-signaling costs are assumed configuration independent and are omitted from the optimization.
\vspace{-5mm}
\subsection{Communication Model}

We assume block fading within each slot and perfect CSI at the BS. All devices transmit simultaneously over bandwidth $B$. The received signal is
$\mathbf r(t)
    =
    \sum_{j\in\mathcal K}
    \sqrt{p_j(t)}\,\mathbf h_j(t)s_j(t)
    +
    \mathbf n(t),$
where
$\mathbf h_j(t)\in\mathbb C^M$ is the uplink channel,
$\mathbb E[|s_j(t)|^2]=1$, and
$\mathbf n(t)\sim\mathcal{CN}
(\mathbf 0,\sigma^2\mathbf I_M)$ is circularly symmetric complex Gaussian noise with covariance $\sigma^2\mathbf I_M$.
For receive beamformer $\mathbf v_k(t)\in\mathbb C^M$, the resulting SINR is
\vspace{-4mm}
\begin{equation}
    \gamma_k(t)
    =
    \frac{
        p_k(t)
        \left|
        \mathbf v_k^H(t)\mathbf h_k(t)
        \right|^2
    }{
        \displaystyle
        \sum_{j\in\mathcal K\setminus\{k\}}
        p_j(t)
        \left|
        \mathbf v_k^H(t)\mathbf h_j(t)
        \right|^2
        +
        \sigma^2\|\mathbf v_k(t)\|^2
    }.
    \label{eq:sinr}
    \vspace{-2mm}
\end{equation}
Assuming Gaussian signaling and capacity-achieving channel coding, the achievable uplink rate is
$    R_k(t)
    =
    B\log_2\big(1+\gamma_k(t)\big)$.

The edge server is assumed to reserve a fixed compute share for each admitted inference request, so that $T_{k,c}^{\rm inf}$ is deterministic and queueing latency is negligible. For the selected configuration, the task latency and communication-plus-inference energy are
\vspace{-2mm}
\begin{align}
    T_k(t)
    &=
    \frac{S_{k,c_k(t)}}{R_k(t)}
    +
    T_{k,c_k(t)}^{\rm inf}
    +
    T_k^{\rm fb},
    \label{eq:total_latency}
    \end{align}
    \vspace{-5mm}
\begin{align}
    E_k(t)
    &=
    p_k(t)\frac{S_{k,c_k(t)}}{R_k(t)}
    +
    P_{k,c_k(t)}^{\rm inf}
    T_{k,c_k(t)}^{\rm inf},
    \label{eq:energy}
\end{align}
where $T_k^{\rm fb}$ is the fixed feedback latency. In \eqref{eq:energy}, the
first term is transmission energy and the second is edge energy attributed to
device $k$.
\vspace{-5mm}
\subsection{Pre-Inference Task-Risk Model}
\label{subsec:task_confidence_model}

Let
$\ell_k:\mathcal A_k\times\mathcal A_k\rightarrow[0,1]$
denote the normalized physical-action loss 
$    \ell_k(\widehat{\mathbf a},\mathbf a)
    =
    \min
    \{
        1,\,
        \frac{
            \|\widehat{\mathbf a}-\mathbf a\|_2^2
        }{
            \delta_k^2
        }
    \}$,
where $\delta_k>0$ is the application-approved action-error scale.
Also, let $c_k^{\rm ref}\in\mathcal C_k$ denote a fixed reference configuration
selected during offline profiling. The reference configuration defines the failure event used to train and calibrate the uncertainty head; online, $q_k^{\rm loc}$ estimates its failure probability directly from $\mathbf z_k(t)$ before $c_k(t)$ is selected. For the random sensory input
$\mathbf X_k$, define
\vspace{-2mm}
{\small \begin{equation}
    \mathbf Z_k
    =
    f_k^{\rm loc}(\mathbf X_k),
    \qquad
    \widehat{\mathbf Y}_k^{\rm ref}
    =
    g_{k,c_k^{\rm ref}}
    (
        e_{k,c_k^{\rm ref}}(\mathbf Z_k)
    ).
    \label{eq:reference_edge_action}
    \vspace{-3mm}
\end{equation}}

For an application-approved failure threshold
$\varepsilon_k\in[0,1)$, the reference action-failure indicator is
\vspace{-2mm}
\begin{equation}
    F_k^{\rm ref}
    =
    \mathbb I
    \{
        \ell_k
        (
            \widehat{\mathbf Y}_k^{\rm ref},
            \mathbf Y_k
        )
        >
        \varepsilon_k
    \}.
    \label{eq:reference_failure_event}
    \vspace{-2mm}
\end{equation}
Thus, $F_k^{\rm ref}=1$ indicates that the action produced by the reference edge-inference configuration is insufficiently accurate for physical execution.
The population pre-inference failure-probability function under the
offline calibration distribution $\mathbb P_k^{\rm cal}$ is
\vspace{-2mm}
\begin{equation}
    q_k^\star(\mathbf z)
    =
    \Pr_{\mathbb P_k^{\rm cal}}
    (
        F_k^{\rm ref}=1
        \,|\,
        \mathbf Z_k=\mathbf z
    ).
    \label{eq:failure_probability_function}
    \vspace{-2mm}
\end{equation}
The deployed head
$q_k^{\rm loc}:\mathbb R^{d_k}\rightarrow[0,1]$
is trained and post-hoc calibrated on a held-out labeled dataset to estimate
$q_k^\star(\cdot)$~\cite{lecoz2024calibration}. Then, the pre-inference uncertainty reported to the BS is
\vspace{-2mm}
\begin{equation}
    u_k(t)=q_k^{\rm loc}(\mathbf z_k(t)).
    \vspace{-2mm}
\end{equation}
Thus, $u_k(t)$ estimates the conditional probability that the fixed reference edge-inference configuration produces an action whose normalized loss exceeds $\varepsilon_k$.

Let $d_k(n)\in[0,d_k^{\max}]$ denote the application-approved physical
consequence assigned to executing an action whose loss exceeds
$\varepsilon_k$ during interval $n$. The normalized physical criticality is
\vspace{-2mm}
\begin{equation}
    \rho_k(n)
    =
    \frac{d_k(n)}{d_k^{\max}}.
    \vspace{-2mm}
    \label{eq:physical_criticality}
    \vspace{-1mm}
\end{equation}

Hence, the normalized pre-transmission task-risk metric is 
\vspace{-2mm}
\begin{equation}
    \xi_k(t)=u_k(t)\rho_k(n).
    \vspace{-2mm}
\end{equation}
Because $u_k(t)$ estimates the reference-configuration failure probability
and $\rho_k(n)$ represents normalized failure consequence, $\xi_k(t)$ determines the required wireless protection.

We propose the adaptive SINR requirement
\vspace{-2mm}
\begin{equation}
    \Gamma_k(t)
    =
    \Gamma_{\min}
    +
    (\Gamma_{\max}-\Gamma_{\min})\xi_k(t),
    \label{eq:adaptive_sinr_requirement}
    \vspace{-2mm}
\end{equation}
where
$0\leq\Gamma_{\min}\leq\Gamma_{\max}$
are linear-scale SINR targets. The affine mapping provides a bounded monotone reliability rule:
higher pre-inference uncertainty or physical criticality increases the
required SINR. 
\vspace{-6mm}
\subsection{Inference-Quality Model}

For a random task input $\mathbf X_k$, the
edge-generated action under configuration $c\in\mathcal C_k$ is
the random variable
$
\widehat{\mathbf Y}_{k,c}
=
g_{k,c}
\left(
e_{k,c}
\big(
f_k^{\rm loc}(\mathbf X_k)
\big)
\right)$.
The slot-level realized action in \eqref{eq:configured_edge_inference} is a
realization of $\widehat{\mathbf Y}_{k,c}$ for
$\mathbf X_k=\mathbf x_k(t)$.

The uncertainty-conditioned expected physical-action loss is
\vspace{-2mm}
\begin{equation}
    \mathcal L_k(c,u)
    =
    \mathbb E_{\mathbb P_k^{\rm cal}}
    [
        \ell_k(\widehat{\mathbf Y}_{k,c},\mathbf Y_k)
        \mid
        U_k=u
    ],
    \qquad
    c\in\mathcal C_k .
    \label{eq:conditional_inference_loss}
    \vspace{-2mm}
\end{equation}
where $ U_k
    =
    q_k^{\rm loc}
    \left(
        f_k^{\rm loc}(\mathbf X_k)
    \right)$ is the random pre-inference uncertainty. For each configuration, $\mathcal L_k(c,u)$ is estimated offline from labeled data and stored as a performance profile. Under the adopted threshold-based digital-link abstraction, feature recovery is assumed successful when the SINR requirement is met, while violations are captured through the reliability constraints and outage metric. The uncertainty head and profiles are assumed valid within the validator-approved deployment domain; out-of-domain tasks use the fallback in Section~\ref{sec:proposed_solution}.

\vspace{-6mm}
\subsection{Risk-aware Joint Design}

At slot $t\in\mathcal T_n$, the BS jointly selects the feature/inference configurations, transmit power levels, and receive beamformers to minimize the profiled expected physical-action loss, end-to-end action latency, and communication plus edge-inference energy as
\vspace{-1mm}
{\small \begin{subequations}
\label{problem:tc_jbei}
\begin{align}
\hspace{-5mm}
    \min_{\{\mathbf v_k(t),p_k(t),c_k(t)\}_{k\in\mathcal K}}
    &
    \sum_{k\in\mathcal K}
    \Big[
        \lambda_{\rm I}
        \mathcal L_k(c_k(t),u_k(t))
        +
        \lambda_{\rm T}T_k(t)
        +
        \lambda_{\rm E}E_k(t)
    \Big]
    \label{problem:tc_jbei_obj}
    \\
    \mathrm{s.t.}\quad&
    \gamma_k(t)\geq\Gamma_k(t),
    \qquad \forall k\in\mathcal K,
    \label{problem:tc_jbei_sinr}
    \\
    &
    T_k(t)\leq T_k^{\max}(n),
    \qquad \forall k\in\mathcal K,
    \label{problem:tc_jbei_latency}
    \\
    &
    0\leq p_k(t)\leq P_k^{\max},
    \qquad \forall k\in\mathcal K,
    \label{problem:tc_jbei_power}
    \\
    &
    c_k(t)\in\mathcal C_k,
    \qquad \forall k\in\mathcal K,
    \label{problem:tc_jbei_config}
    \\
    &
    \|\mathbf v_k(t)\|^2=1,
    \qquad \forall k\in\mathcal K.
    \label{problem:tc_jbei_bf}
    \vspace{-6mm}
\end{align}
\end{subequations}}
The weights $\lambda_{\rm I}$, $\lambda_{\rm T}$, and $\lambda_{\rm E}$ put inference loss, latency, and energy on a common cost scale. The constraints enforce risk-aware reliability, end-to-end deadline satisfaction, transmit-power limits, profiled configuration selection, and unit-norm receive beamforming.

Problem~\eqref{problem:tc_jbei} is mixed discrete--continuous and nonconvex because of the discrete configuration variables, coupled multiuser SINRs, and nonlinear rate-dependent latency and energy. Unlike fixed quality-of-service (QoS) formulations, the required wireless reliability depends on the estimated task-risk, while the inference cost depends on both the current uncertainty and selected configuration.
\vspace{-4mm}
\section{Proposed Solution}
\label{sec:proposed_solution}

We propose RACE to approximate repeated slot-level solutions of~\eqref{problem:tc_jbei} by separating slow task-contract validation from fast wireless control. A hybrid actor selects discrete inference configurations and continuous initial transmit power levels, while a model-based layer computes receive beamformers and performs finite power restoration. The LLM only proposes contract parameters, which are validated before use, and does not make time-critical radio decisions. This follows the separation of intent interpretation and numerical control in LLM-enabled network management~\cite{cai2025llmrlwireless, zheng2026llmrlwireless}.
\vspace{-5mm}
\subsection{Slow-Timescale Task-Service Contract Validation}

Let $\mathcal D_k(n)$ denote the structured description
of the physical task performed by device $k$ during task interval $n$. The
LLM first generates the candidate contract
\vspace{-2mm}
\begin{equation}
\widehat{\boldsymbol{\tau}}_k(n)
=
F_{\rm LLM}
\left(
\mathcal D_k(n),
\mathbf e_k(n-1)
\right),
\label{eq:candidate_task_contract}
\vspace{-2mm}
\end{equation}
where $\mathbf e_k(n-1)$ is an observable summary of the preceding interval, such as the average latency and energy and the measured SINR- and
deadline-violation rates. The candidate contract contains a proposed physical-criticality level and action deadline, $\widehat{\boldsymbol{\tau}}_k(n) =[\widehat{\rho}_k(n),\widehat{T}_k^{\max}(n)]^T$. The summary does not contain the true physical-action loss, since the ideal action $\mathbf Y_k(t)$ is unavailable during online operation.

A deterministic validator then produces the deployed contract as
\vspace{-2mm}
\begin{equation}
\boldsymbol{\tau}_k(n)
=
\mathcal V_k
\left(
\widehat{\boldsymbol{\tau}}_k(n),
\mathcal D_k(n)
\right)
=
\left[
\rho_k(n),
T_k^{\max}(n)
\right]^T.
\label{eq:validated_task_contract}
\vspace{-2mm}
\end{equation}
The validator parses the output, checks that $\mathcal D_k(n)$ lies within the calibrated deployment domain, and maps the proposed values to approved criticality levels $\rho_k(n)\in\mathcal R_k\subseteq[0,1]$ and deadlines $T_k^{\max}(n)\in\mathcal Q_k$. Here, $\mathcal R_k$ comprises normalized, application-approved failure consequences rather than unrestricted LLM scores, while $\mathcal Q_k$ contains approved deadlines. 

The validator also checks static deadline feasibility. Define the
transmission-time budget for configuration $c\in\mathcal C_k$ as
\vspace{-2mm}
\begin{equation}
\Delta T_{k,c}(n)
=
T_k^{\max}(n)
-
T_{k,c}^{\rm inf}
-
T_k^{\rm fb}.
\label{eq:transmission_time_budget}
\vspace{-2mm}
\end{equation}
A candidate contract is statically admissible only if
$\Delta T_{k,c}(n)>0$ for at least one $c\in\mathcal C_k$. Malformed,
out-of-domain, or statically infeasible candidate contracts are replaced by
the most recent valid contract; a predefined application-approved safe
contract is used for the initial interval.

The validated contract remains fixed over all
$t\in\mathcal T_n$. Hence, the LLM and validator operate only at the task
timescale and are excluded from the slot-level wireless control loop. Within
the validator-approved deployment domain, the offline profiles
$\mathcal L_k(c,u)$ are assumed to remain valid conditional surrogates for
comparing the profiled configurations.
\vspace{-5mm}
\subsection{Contract-Induced Effective SINR Requirement}

For a configuration $c\in\mathcal C_k$ satisfying
$\Delta T_{k,c}(n)>0$, the latency constraint
\eqref{problem:tc_jbei_latency} is equivalent to
\vspace{-3mm}
\begin{equation}
\gamma_k(t)
\geq
\Gamma_{k,c}^{\rm lat}(n)
\triangleq
2^{
\frac{
S_{k,c}
}{
B\Delta T_{k,c}(n)
}
}
-1.
\label{eq:latency_induced_sinr}
\vspace{-2mm}
\end{equation}
Combining this condition with the risk-aware SINR requirement
\eqref{problem:tc_jbei_sinr} gives the effective target
\vspace{-3mm}
\begin{equation}
\overline{\Gamma}_{k,c}(t;n)
=
\max
\{
\Gamma_k(t),
\Gamma_{k,c}^{\rm lat}(n)
\}.
\label{eq:effective_sinr_target}
\vspace{-2mm}
\end{equation}
Therefore, for the selected configuration $c_k(t)$,
$\gamma_k(t)\geq
\overline{\Gamma}_{k,c_k(t)}(t;n)$
jointly enforces the SINR and latency constraints.

Before categorical action sampling, the BS constructs the state-dependent
admissible configuration set
$\mathcal C_k^{\rm adm}(t,n)
=
\{
c\in\mathcal C_k \;|\;
\Delta T_{k,c}(n)>0,\
\overline{\Gamma}_{k,c}(t;n)
\leq
\frac{
P_k^{\max}\|\mathbf h_k(t)\|^2
}{
\sigma^2
}
\}$.

The second condition is a necessary interference-free feasibility test that masks configurations unable to meet the effective SINR target even at maximum power. If $\mathcal C_k^{\rm adm}(t,n)=\varnothing$ for any device, the slot is declared infeasible and the application-defined fallback is invoked. Otherwise, residual multiuser infeasibility is handled by restoration and, if unresolved, outage declaration.
\vspace{-5mm}
\subsection{Fast-Timescale Contextual Hybrid Actor--Critic}

At slot $t\in\mathcal T_n$, the BS observes the channel matrix
$\mathbf H(t)=[\mathbf h_1(t),\ldots,\mathbf h_K(t)]$, the pre-inference
uncertainty vector
$\mathbf u(t)=[u_1(t),\ldots,u_K(t)]^T$, and the validated contract vector
$\boldsymbol{\tau}(n)=[\boldsymbol{\tau}_1^T(n),\ldots,
\boldsymbol{\tau}_K^T(n)]^T$.
The controller state is
$\mathbf s(t)
=
[
\operatorname{vec}\!\left(\Re\{\mathbf H(t)\}\right)^T,
\operatorname{vec}\!\left(\Im\{\mathbf H(t)\}\right)^T,
\mathbf u^T(t),
\boldsymbol{\tau}^T(n)
]^T$,
where $\operatorname{vec}(\cdot)$ denotes the column-wise vectorization
operator. 
The actor outputs the raw hybrid action
$\widetilde{\mathbf a}(t)=(\mathbf c(t),\widetilde{\mathbf p}(t))$, where
$\mathbf c(t)=[c_1(t),\ldots,c_K(t)]^T$ and
$\widetilde{\mathbf p}(t)=[\widetilde p_1(t),\ldots,\widetilde p_K(t)]^T$.
The joint policy is factorized as
\vspace{-2mm}
{\small\begin{align}
\pi_{\boldsymbol\theta}
(\mathbf c,\widetilde{\mathbf p}\mid\mathbf s)
&=
\pi_{\boldsymbol\theta}^{c}(\mathbf c\mid\mathbf s)
\pi_{\boldsymbol\theta}^{p}
(\widetilde{\mathbf p}\mid\mathbf s,\mathbf c),
\nonumber\\
\pi_{\boldsymbol\theta}^{c}(\mathbf c\mid\mathbf s)
&=
\prod_{k\in\mathcal K}
\pi_{\boldsymbol\theta,k}^{c}
(c_k\mid\mathbf s,\mathcal C_k^{\rm adm}(t,n)),
\nonumber\\
\pi_{\boldsymbol\theta}^{p}
(\widetilde{\mathbf p}\mid\mathbf s,\mathbf c)
&=
\prod_{k\in\mathcal K}
\pi_{\boldsymbol\theta,k}^{p}
(\widetilde p_k\mid\mathbf s,\mathbf c).
\label{eq:hybrid_policy_factorization}
\vspace{-5mm}
\end{align}}
where each categorical branch is renormalized over the corresponding
admissible set. Conditioning the continuous branch on $\mathbf c$ allows the
power policy to account explicitly for the configuration-dependent feature
sizes, inference times, energy profiles, and effective SINR targets.

For an admitted inference request, the raw power is 
$\widetilde p_k(t)
=
p_k^{\min}
+
\left(
P_k^{\max}-p_k^{\min}
\right)
X_k(t)$
,
where $X_k(t)\sim{\rm Beta}(\alpha_k(t),\beta_k(t))$, $\alpha_k(t),\beta_k(t)>0$ are actor outputs, and $p_k^{\min}>0$ is a small numerical implementation floor. This avoids the zero-rate singularity in \eqref{eq:total_latency} and \eqref{eq:energy}; for admitted requests with positive feature size and finite deadline, feasible transmission already requires strictly positive power.

Since \eqref{problem:tc_jbei} is slot-wise and has no queue, battery, or action-dependent physical-state transition, RACE is trained as a contextual actor--critic. Each slot constitutes a contextual decision, and the critic $V_{\boldsymbol\phi}(\mathbf s(t))$ estimates the expected immediate reward under the current policy, providing a variance-reduction baseline.
\vspace{-6mm}
\subsection{Model-Based Beamforming and Feasibility Restoration}

For any power vector $\mathbf p(t)$, define the received covariance matrix
\vspace{-2mm}
\begin{equation}
\mathbf C(t)
=
\sum_{j\in\mathcal K}
p_j(t)
\mathbf h_j(t)\mathbf h_j^H(t)
+
\sigma^2\mathbf I_M.
\label{eq:received_covariance}
\end{equation}

The SINR-maximizing unit-norm linear receiver is
\vspace{-2mm}
\begin{equation}
\mathbf v_k(\mathbf p(t))
=
\frac{\mathbf C^{-1}(t)\mathbf h_k(t)}
{\|\mathbf C^{-1}(t)\mathbf h_k(t)\|}.
\label{eq:model_based_beamformer}
\end{equation}
Equation~\eqref{eq:model_based_beamformer} is directionally equivalent to the interference-plus-noise inverse receiver obtained by excluding the desired rank-one term from $\mathbf C(t)$. The shared covariance matrix requires only one matrix factorization per restoration iteration for all devices.

Because categorical masking alone does not guarantee simultaneous
multiuser feasibility, the actor-selected power levels are passed through a
lightweight model-based restoration layer. Initialize
$\mathbf p^{(0)}(t)
=
\widetilde{\mathbf p}(t)$.
At restoration iteration $q$, the BS computes
$\mathbf v_k^{(q)}(t)
=
\mathbf v_k(\mathbf p^{(q)}(t))$ using \eqref{eq:model_based_beamformer} and evaluates
the power required to meet the effective target under the current interference level
\vspace{-1mm}
{\small\begin{equation}
\psi_k^{(q)}(t)
=
\overline{\Gamma}_{k,c_k(t)}(t;n)
\frac{
\displaystyle
\sum_{j\in\mathcal K\setminus\{k\}}
p_j^{(q)}(t)
|
(\mathbf v_k^{(q)}(t))^H
\mathbf h_j(t)
|^2
+
\sigma^2
\|
\mathbf v_k^{(q)}(t)
\|^2
}{
|
(\mathbf v_k^{(q)}(t))^H
\mathbf h_k(t)
|^2
}.
\label{eq:required_power_update}
\vspace{-2mm}
\end{equation}
}
The power update is
\vspace{-2mm}
\begin{equation}
p_k^{(q+1)}(t)
=
\min
\{
P_k^{\max},
\max
\{
p_k^{(q)}(t),
\psi_k^{(q)}(t)
\}
\}.
\label{eq:power_restoration}
\vspace{-2mm}
\end{equation}
Thus, the restoration layer preserves actor-selected power levels that are already
sufficient and only increases power levels that violate the effective SINR
requirements. The updates are repeated for at most $Q_{\rm R}$ iterations or
until
\vspace{-2mm}
\begin{equation}
\gamma_k^{(q)}(t)
\geq
\overline{\Gamma}_{k,c_k(t)}(t;n)-\epsilon_{\rm R},
\qquad
\forall k\in\mathcal K,
\label{eq:restoration_stopping_rule}
\vspace{-2mm}
\end{equation}
where $\epsilon_{\rm R}\geq0$ is a numerical tolerance.
Here, $\gamma_k^{(q)}(t)$ denotes the SINR obtained from
$\mathbf p^{(q)}(t)$ and $\mathbf v_k^{(q)}(t)$.

After each power update, the receive beamformers and SINRs are
recomputed; let $Q_t\leq Q_{\rm R}$ denote the final evaluated iterate. The executed decisions are
$\mathbf p^\star(t)=\mathbf p^{(Q_t)}(t)$ and
$\mathbf v_k^\star(t)=\mathbf v_k^{(Q_t)}(t)$. 
Define the post-restoration feasibility indicator
\vspace{-2mm}
\begin{equation}
\zeta(t)
=
\mathbb I
\left\{
\gamma_k^\star(t)
\geq
\overline{\Gamma}_{k,c_k(t)}(t;n)-\epsilon_{\rm R},
\quad
\forall k\in\mathcal K
\right\},
\label{eq:feasibility_indicator}
\vspace{-2mm}
\end{equation}
where $\gamma_k^\star(t)$ is the SINR obtained from the executed decisions
$\mathbf p^\star(t)$ and $\mathbf v_k^\star(t)$. When $\zeta(t)=1$, the executed action satisfies the SINR and latency
constraints up to the prescribed numerical tolerance. When $\zeta(t)=0$, the selected configurations and actor-generated
initial power levels are not restored to feasibility by the adopted finite-iteration
procedure; the slot is declared an outage and its residual violations are
penalized during training.
\vspace{-2mm}
\begin{prop}[Bounded Restoration and Conditional Feasibility]
\label{prop:restoration_feasibility}
Consider slot $t\in\mathcal T_n$ and a configuration vector
$\mathbf c(t)$ selected from the admissible configuration sets
$\{\mathcal C_k^{\rm adm}(t,n)\}_{k\in\mathcal K}$. Assume that
$\mathbf h_k(t)\neq\mathbf 0$ for all $k\in\mathcal K$ and that
$0\leq p_k^{(0)}(t)\leq P_k^{\max}$ for all $k\in\mathcal K$. Then, the
power sequence generated by \eqref{eq:power_restoration} satisfies
\vspace{-2mm}
\begin{equation}
0\leq p_k^{(q)}(t)\leq P_k^{\max},
\quad
p_k^{(q+1)}(t)\geq p_k^{(q)}(t),
\quad
\forall k\in\mathcal K,\ \forall q.
\label{eq:restoration_monotonicity}
\vspace{-2mm}
\end{equation}
Hence, $\{\mathbf p^{(q)}(t)\}$ is component-wise convergent. Moreover, if
the finite restoration terminates with $\zeta(t)=1$, then
$0\leq p_k^\star(t)\leq P_k^{\max}$ and
$\|\mathbf v_k^\star(t)\|^2=1$ for all $k$, and
\vspace{-2mm}
\begin{equation}
\gamma_k^\star(t)
\geq
\overline{\Gamma}_{k,c_k(t)}(t;n)-\epsilon_{\rm R},
\qquad \forall k\in\mathcal K.
\label{eq:restored_effective_sinr}
\vspace{-2mm}
\end{equation}
Consequently, for $\epsilon_{\rm R}=0$, the executed action satisfies the
power, beamformer-normalization, risk-aware SINR, and task-latency
constraints in
\eqref{problem:tc_jbei_power},
\eqref{problem:tc_jbei_bf},
\eqref{problem:tc_jbei_sinr}, and
\eqref{problem:tc_jbei_latency}. For $\epsilon_{\rm R}>0$, the executed
action satisfies the effective-SINR condition up to the prescribed
linear-SINR tolerance.
\end{prop}
\vspace{-2mm}
\begin{IEEEproof} See Appendix~\ref{app:proof_prop}. \end{IEEEproof}
\vspace{-4mm}
\subsection{Constraint-Aware Reward and PPO Training}

Let $\mathcal J(t)$ denote the objective in
\eqref{problem:tc_jbei_obj} evaluated using the executed decisions
$\mathbf c(t)$, $\mathbf p^\star(t)$, and
$\{\mathbf v_k^\star(t)\}_{k\in\mathcal K}$. Define the normalized residual effective-SINR violation as
$\delta_k^\Gamma(t)=
[1-\gamma_k^\star(t)/\overline{\Gamma}_{k,c_k(t)}(t;n)]^+$.
The constraint-aware reward is
\vspace{-1mm}
{\small \begin{align}
\hspace{-6.2mm}
r(t)
=
{}&
-\mathcal J(t)
-
\lambda_\Gamma
\sum_{k\in\mathcal K}
\delta_k^\Gamma(t)
-
\lambda_{\rm R}
\sum_{k\in\mathcal K}
(
\frac{
p_k^\star(t)-\widetilde p_k(t)
}{
P_k^{\max}
}
)^2 -
\lambda_{\rm O}
\big(1-\zeta(t)\big),
\label{eq:constraint_aware_reward}
\vspace{-3mm}
\end{align}
}
where
$\lambda_\Gamma,\lambda_{\rm R},\lambda_{\rm O}>0$.
The reward minimizes the risk-aware objective, penalizes residual SINR
violation, discourages reliance on restoration, and penalizes unrestorable
outage states.

For a contextual sample at slot $t$, the hybrid-policy likelihood ratio is
\vspace{-1mm}
\begin{equation}
\varrho_t(\boldsymbol\theta)
=
\frac{
\pi_{\boldsymbol\theta}
(\mathbf c(t),\widetilde{\mathbf p}(t)\mid\mathbf s(t))
}{
\pi_{\boldsymbol\theta_{\rm old}}
(\mathbf c(t),\widetilde{\mathbf p}(t)\mid\mathbf s(t))
}.
\label{eq:hybrid_ppo_ratio}
\vspace{-2mm}
\end{equation}
Using the one-step advantage estimate
$\widehat A(t)
=
r(t)
-
V_{\boldsymbol\phi}
\left(
\mathbf s(t)
\right)$,
the actor is updated by maximizing the clipped proximal policy optimization (PPO) surrogate
\vspace{-1mm}
{\small \begin{align}
\mathcal L_{\rm PPO}(\boldsymbol\theta)
=
\mathbb E_t
\big[
\min
\{
\varrho_t(\boldsymbol\theta)\widehat A(t),
\operatorname{clip}
(
\varrho_t(\boldsymbol\theta),
1-\epsilon_{\rm PPO},
1+\epsilon_{\rm PPO}
)
\widehat A(t)
\}
\big].
\label{eq:ppo_objective}
\vspace{-6mm}
\end{align}}
while the critic minimizes
$\mathcal L_{\rm V}(\boldsymbol\phi)
=
\mathbb E_t
[
(
V_{\boldsymbol\phi}(\mathbf s(t))
-
r(t)
)^2
]$.
The actor and critic are centrally trained at the BS using the joint probability of the masked categorical and bounded continuous action branches \cite{schulman2017proximal}. During execution, the learned actor provides an amortized approximation to the repeated mixed discrete--continuous problem, while the analytical receiver and restoration layer exploit wireless structure to improve feasibility and reduce the learning burden.
\vspace{-3mm}
\section{Simulation Results} \label{sec:simulation_results}
We evaluate RACE in the interference-coupled uplink setting specified in Table~\ref{tab:simulation_parameters}. The slow-timescale LLM--validator is represented by its validated task-contract outputs, $\boldsymbol\tau_k(n)$. All methods share the same analytical receive beamforming rule, enabling consistent comparison of inference configurations, power control, and risk-dependent resource allocation.

\begin{table}[t]
\vspace*{-2mm}
\centering
\caption{Simulation Parameters} \vspace{-3mm} \label{tab:simulation_parameters} 
\scriptsize \setlength{\tabcolsep}{3pt} \renewcommand{\arraystretch}{0.9} \begin{tabular}{l|l}
\hline \textbf{Parameter} & \textbf{Value} \\ \hline Network & $K=6$, $M=12$, $B=1$ MHz, $\sigma^2=10^{-3}$ \\ Channel & $\mathbf h_k(t)=\sqrt{\beta_k}\mathbf g_k(t)$, $\mathbf g_k(t)\sim\mathcal{CN}(\mathbf0,\mathbf I_M)$ \\ Large-scale factor & $\beta_k\sim{\rm Unif}[0.02,0.20]$ \\ Power/SINR & $P_k^{\max}=0.24$ W, $p_k^{\min}=10^{-5}$ W, $\Gamma_{\min}=-5$ dB, $\Gamma_{\max}=10$ dB \\ Task contract & $\rho_k(n)\in\{0.10,0.30,0.60,0.90\}$, $T_k^{\max}(n)\in\{70,105,150,220\}$ ms \\ Uncertainty & $u_k\sim0.65{\rm Beta}(2,5)+0.35{\rm Beta}(5,2)$ \\ Profiles & $S_{k,c}=[4,8,16,40,80]$ kbits, $T_k^{\rm fb}=5$ ms \\ Inference & $T_{k,c}^{\rm inf}=[6,14,30,88,175]$ ms, $P_{k,c}^{\rm inf}=[0.07,0.14,0.27,0.72,1.15]$ W \\ Weights/restoration & $\lambda_{\rm I}=2$, $\lambda_{\rm T}=3$, $\lambda_{\rm E}=1$, $Q_{\rm R}=8$, $\epsilon_{\rm R}=0$ \\ Evaluation & $5$ random seeds, $2000$ slots/seed \\ \hline \end{tabular}\vspace{-2mm}
\end{table}

We compare RACE with seven baselines. Fixed-high
and Fixed-low
replace SINR requirement $\Gamma_k(t)$ with $\Gamma_{\max}$ and $\Gamma_{\min}$, respectively, while retaining $\Gamma_{k,c}^{\rm lat}(n)$. Risk-greedy employs a one-step risk-aware heuristic; Uncertainty-only removes physical criticality by setting $\xi_k(t)=u_k(t)$; No-contract uses average task contracts for control but is evaluated against the actual task-specific requirements; Fixed-config uses the middle admissible inference configuration; and No-restoration disables power restoration. The reported accuracy--latency--energy cost excludes training penalties, while constraint violations are evaluated separately through outage and residual effective-SINR violation. 

\begin{figure}[!t] 
\vspace*{-2mm}
\centering
\includegraphics[width=0.6\linewidth]{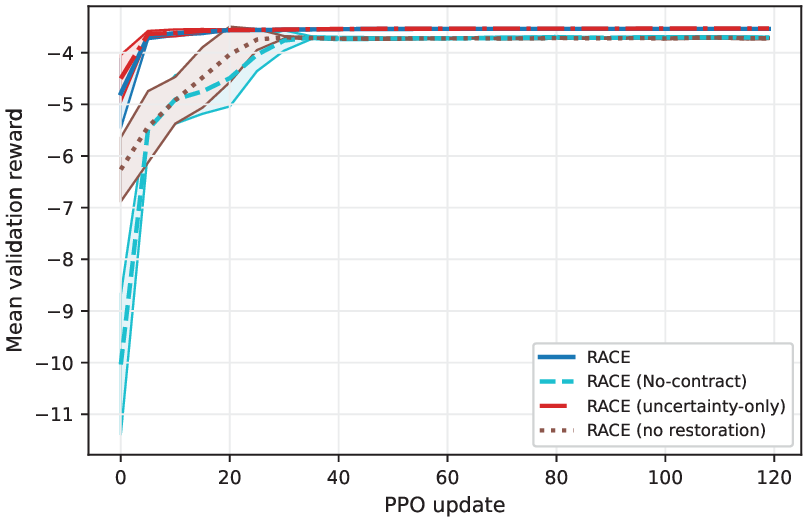}
\vspace*{-4mm}
\caption{Training convergence behavior.}
\vspace{-5mm}
\label{fig:training_convergence}
\end{figure}

\begin{figure}[!t]
    \centering
    \includegraphics[width=\linewidth]{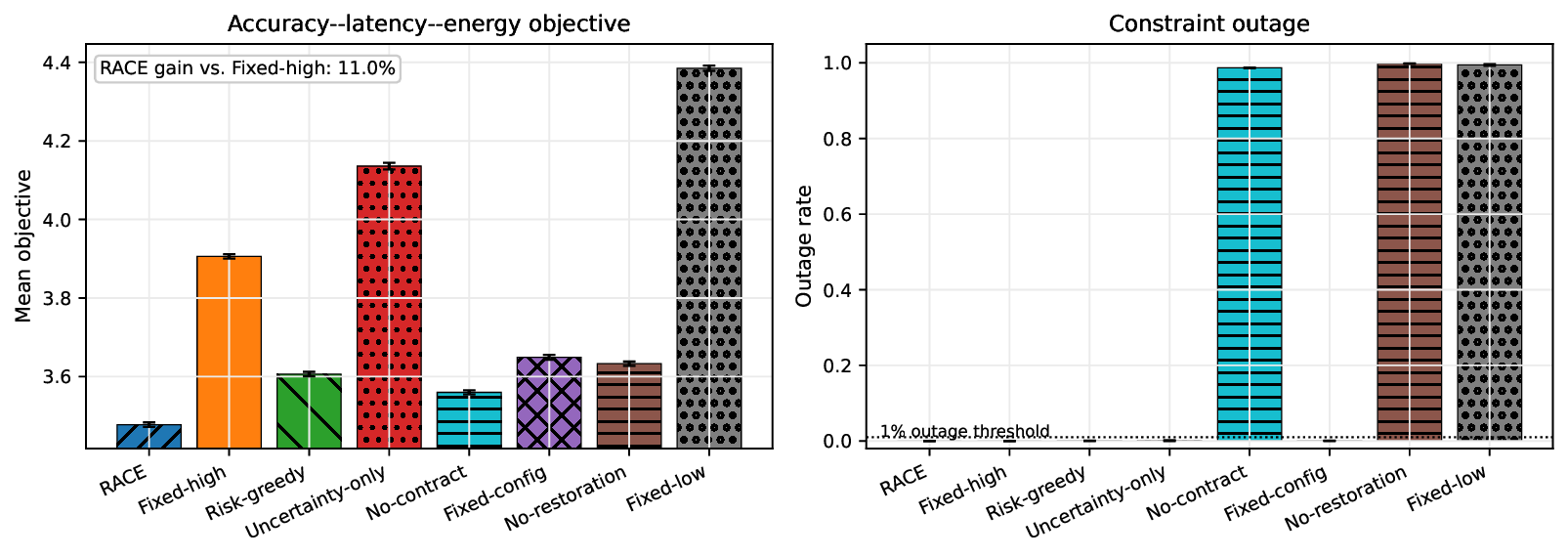}
    \vspace*{-8mm}
    \caption{Average accuracy--latency--energy cost and constraint-outage rate.} 
    \vspace{-5mm}
    \label{fig:objective_outage}
\end{figure}

Fig.~\ref{fig:training_convergence} shows rapid improvement and subsequent stabilization of RACE's validation penalized reward. Fig.~\ref{fig:objective_outage} demonstrates the advantage of risk-aware joint inference and wireless resource control. RACE achieves the lowest accuracy--latency--energy cost among the considered methods, outperforming the reliable Fixed-high baseline by $11\%$ while maintaining near-zero constraint outage. Fixed-high incurs higher cost due to conservative SINR requirements, whereas Fixed-low exhibits higher outage from insufficient risk-dependent SINR provisioning. No-contract and No-restoration also suffer severe outage, demonstrating the importance of task-specific requirements and corrective power restoration for constraint satisfaction. 

\begin{figure}[!t]
    \centering
    \includegraphics[width=\linewidth]{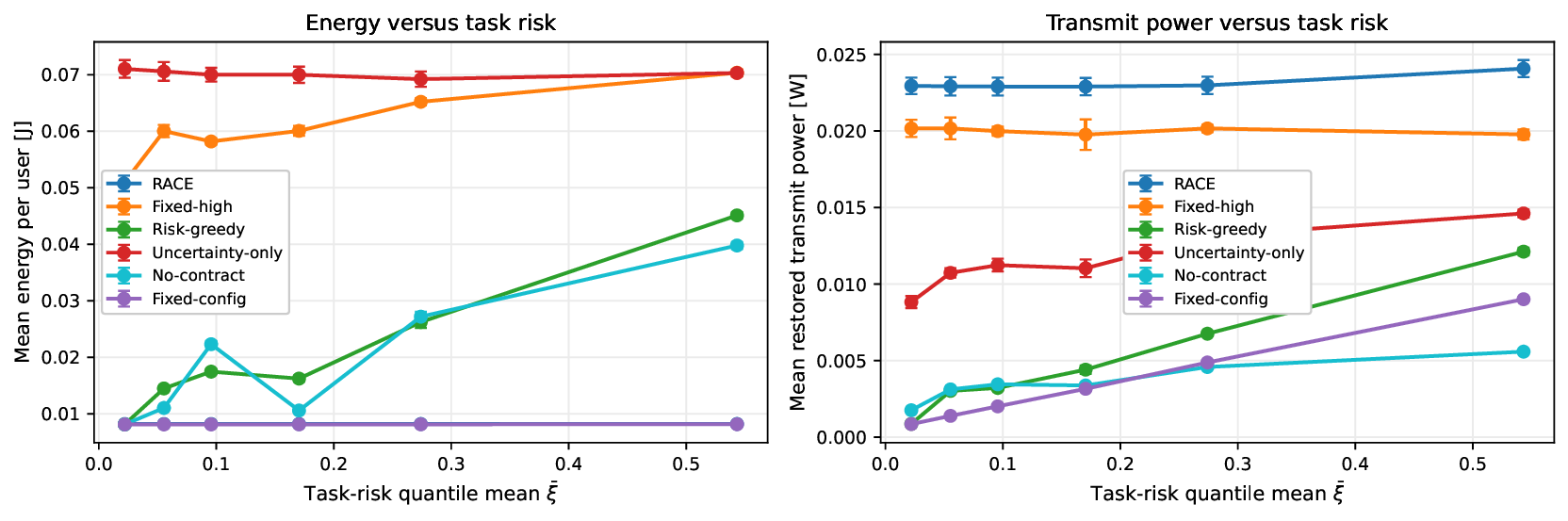}
    \vspace*{-7mm}
    \caption{Energy and restored transmit power versus task-risk quantile.}
    \vspace{-3mm}
    \label{fig:risk_energy_power}
\end{figure}

\begin{figure}[!t]
    \centering
    \includegraphics[width=\linewidth]{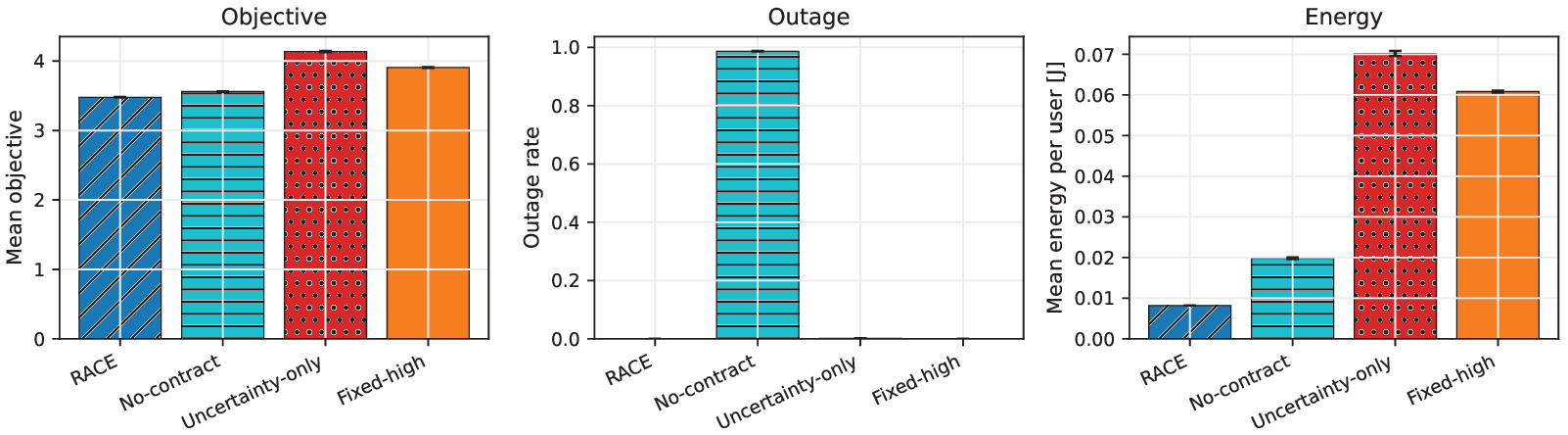}
    \vspace*{-8mm}
    \caption{Slow-timescale contract ablation.}
    \vspace{-2mm}
    \label{fig:contract_ablation}
\end{figure}

Fig.~\ref{fig:risk_energy_power} further reveals how RACE exploits physical-action risk for efficient resource allocation. Across task-risk quantiles, RACE maintains lower energy consumption than Fixed-high and Uncertainty-only through joint adaptation of inference configurations and transmit power levels. The restored transmit-power results show that RACE balances energy efficiency with satisfaction of the effective SINR requirements rather than simply minimizing transmission power. Importantly, the comparison with Uncertainty-only demonstrates the additional benefit of incorporating physical-action consequence beyond inference uncertainty alone. Fig.~\ref{fig:contract_ablation} further demonstrates the role of task-specific contracts: replacing the validated criticality--deadline pair with average values leads to severe constraint outage, whereas RACE maintains near-zero outage and the lowest cost among reliable contract-aware methods.


\vspace{-4mm}
\section{Conclusion} \label{sec:conclusion} This letter proposed RACE, a risk-aware framework for joint edge-inference and wireless resource control in Physical-AI networks. RACE maps predicted action-failure probability and validated physical criticality to adaptive SINR targets, combines them with deadline-derived requirements, and employs hybrid actor--critic control with analytical receive beamforming and finite power restoration. Simulations demonstrated an 11\% reduction in average accuracy--latency--energy cost relative to fixed-high-SINR control, with near-zero observed constraint outage.
\vspace{-3mm}
\appendices
\renewcommand{\theequation}{\thesection.\arabic{equation}}
\section{Proof of Proposition~\ref{prop:restoration_feasibility}}
\label{app:proof_prop}
\setcounter{equation}{0}

\begin{IEEEproof}
The result follows by induction. By assumption,
$0\leq p_k^{(0)}(t)\leq P_k^{\max}$. If
$0\leq p_k^{(q)}(t)\leq P_k^{\max}$, then
\eqref{eq:power_restoration} gives
$p_k^{(q+1)}(t)
=
\min\{P_k^{\max},
\max\{p_k^{(q)}(t),\psi_k^{(q)}(t)\}\}$.
Since $\psi_k^{(q)}(t)\geq0$ and $p_k^{(q)}(t)\leq P_k^{\max}$,
we have
$0\leq p_k^{(q+1)}(t)\leq P_k^{\max}$ and
$p_k^{(q+1)}(t)\geq p_k^{(q)}(t)$. Hence,
$\{p_k^{(q)}(t)\}_q$ is monotone nondecreasing and upper bounded by
$P_k^{\max}$, and is therefore convergent.

At termination,
$\mathbf p^\star(t)=\mathbf p^{(Q_t)}(t)$ for some
$Q_t\leq Q_{\rm R}$, so
$0\leq p_k^\star(t)\leq P_k^{\max}$.
Since $\sigma^2>0$, $\mathbf C(t)$ in
\eqref{eq:received_covariance} is positive definite. Together with
$\mathbf h_k(t)\neq\mathbf0$, this implies
$\mathbf C^{-1}(t)\mathbf h_k(t)\neq\mathbf0$; hence,
\eqref{eq:model_based_beamformer} is well defined and
$\|\mathbf v_k^\star(t)\|^2=1$.

If $\zeta(t)=1$, \eqref{eq:feasibility_indicator} gives
$\gamma_k^\star(t)
\geq
\overline{\Gamma}_{k,c_k(t)}(t;n)-\epsilon_{\rm R}$.
For $\epsilon_{\rm R}=0$,
$\overline{\Gamma}_{k,c_k(t)}(t;n)
=
\max\{\Gamma_k(t),\Gamma_{k,c_k(t)}^{\rm lat}(n)\}$,
so both the risk-aware and latency-induced SINR conditions hold.
By \eqref{eq:latency_induced_sinr}, the latter is equivalent to the
deadline constraint. Thus,
\eqref{problem:tc_jbei_power},
\eqref{problem:tc_jbei_bf},
\eqref{problem:tc_jbei_sinr}, and
\eqref{problem:tc_jbei_latency} hold. For $\epsilon_{\rm R}>0$, the same
conclusion holds up to the prescribed linear-SINR tolerance.
\end{IEEEproof}

\end{document}